\documentclass[11pt,a4paper]{article}

\usepackage{a4wide}
\usepackage{graphicx}
\usepackage{hyperref}

\newcommand{\doi}[1]{\href{http://dx.doi.org/#1}{doi:#1}}
\newcommand{\identifiers}[2]{\href{http://identifiers.org/#1/#2}{#2}}
\newcommand{\tabref}[1]{Table~\ref{table:#1}}
\newcommand{\sbo}[1]{\href{http://www.ebi.ac.uk/sbo/main/SBO:0000#1}{#1}}

\begin{document}

\setlength{\parindent}{0pt}
\addtolength{\parskip}{6pt}
\newlength{\figWidth} \setlength{\figWidth}{0.5\textwidth}

\author{Kieran Smallbone \\ [24pt]
\emph{Manchester Centre for Integrative Systems Biology} \\
\emph{131 Princess Street, Manchester M1 7DN, UK} \\
\href{mailto:kieran.smallbone@manchester.ac.uk}{\tt kieran.smallbone@manchester.ac.uk}
}

\title{Standardized network reconstruction of E.~coli metabolism}

\date{}

\maketitle

\begin{abstract}

\noindent We have created a genome-scale network reconstruction of Escherichia coli metabolism. Existing reconstructions were improved in terms of annotation standards, to facilitate their subsequent use in dynamic modelling. The resultant network is available from EcoliNet~(\href{http://ecoli.sf.net/}{http://ecoli.sf.net/}).

\end{abstract}

\section*{EcoliNet}

The structure of metabolic networks can be determined by a reconstruction approach, using data from genome annotation, metabolic databases and chemical databases~\cite{palsson10}. We built upon an existing reconstruction of the metabolic network of E.~coli that was based on genomic and literature data (known as iJO1366, \cite{orth11}). This model contains 1366 genes, 2251 metabolic reactions, and 1136 unique metabolites. Comparison to experimental data sets shows that it makes accurate phenotypic predictions of growth on different substrates and for gene knockout strains~\cite{orth11}. 

iJO1366 suffers from the use of non-standard names and is not annotated with methods that are machine-readable. The model was thus updated according to existing community-driven annotation standards~\cite{herrgard08}. The reconstruction is described and made available in Systems Biology Markup Language (SBML) (\href{http://sbml.org/}{http://sbml.org/}, \cite{hucka03}), an established community XML format for the mark-up of biochemical models that is understood by a large number of software applications.  The network is available from EcoliNet~(\href{http://ecoli.sf.net/}{http://ecoli.sf.net/}).

\subsection*{Annotation}

The highly-annotated network is primarily assembled and provided as an SBML file. Specific model entities, such as species or reactions, are annotated using ontological terms. These annotations, encoded using the resource description framework (RDF)~\cite{wang05}, provide the facility to assign definitive terms to individual components, allowing  software to identify such components unambiguously and thus link model components to existing data resources~\cite{kell08}. Minimum Information Requested in the Annotation of Models (MIRIAM, \cite{lenovere05}) --compliant annotations have been used to identify components unambiguously by associating them with one or more terms from publicly available databases registered in MIRIAM resources~\cite{laibe08}. Thus this network is entirely traceable and is presented in a computational framework.

Nine different databases are used to annotate entities in the network (see \tabref{1}). The Systems Biology Ontology (SBO)~\cite{courtot11} is also used to semantically discriminate between entity types. Eight different SBO terms are used to annotate entities in the network (see \tabref{2}).

\begin{table}[!ht]
\begin{center}
\begin{tabular}{| c | c | c |}
	\hline 
	example 										& 	identifier				 			& 	database 			\\
	\hline 
	EcoliNet										&	\identifiers{taxonomy}{562}			&	taxonomy 			\\
	EcoliNet										&	\identifiers{pubmed}{21988831}		&	pubmed 			\\
	cytoplasm 									&	\identifiers{obo.go}{GO:0005737}		&	obo.go			\\
	(-)-ureidoglycolate								&	\identifiers{kegg.compound}{C00603}	&	kegg.compound	\\
	(-)-ureidoglycolate								&	\identifiers{chebi}{CHEBI:57296}		&	chebi			\\
	glgB											&	\identifiers{kegg.genes}{eco:b3432}		&	kegg.genes		\\
	glgB											&	\identifiers{uniprot}{P07762}			&	uniprot			\\	
	1,4-alpha-glucan branching enzyme				&	\identifiers{ec-code}{2.4.1.18}			&	ec-code			\\
	2-dehydro-3-deoxygalactonokinase					&	\identifiers{isbn}{1555810845}			&	isbn				\\
	\hline
\end{tabular}
\caption{MIRIAM annotations used in the model.}
\label{table:1}
\end{center}
\end{table}

\begin{table}[!ht]
\begin{center}
\begin{tabular}{| c | c | c |}
	\hline 
	example 										& 	SBO term 		& 	interpretation			\\
	\hline 
	cytoplasm										&	\sbo{290} 		&	compartment			\\
	(-)-ureidoglycolate								&	\sbo{247} 		&	metabolite			\\
	tRNA (Glu)									&	\sbo{250} 		&	ribonucleic acid		\\
	glgB											&	\sbo{252} 		&	enzyme				\\
	1,4-alpha-glucan branching enzyme				&	\sbo{176} 		&	biochemical reaction	\\
	1,4-alpha-glucan transport						&	\sbo{185} 		&	transport reaction		\\	
	biomass objective function						&	\sbo{397} 		&	modelling reaction		\\	
	glgB $\rightarrow$ 1,4-alpha-glucan branching enzyme		&	\sbo{460} 		&	catalyst				\\
	\hline
\end{tabular}
\caption{SBO terms used in the model.}
\label{table:2}
\end{center}
\end{table}

\subsection*{Use}

We maintain the distinction between the E.~coli GEnome scale Network REconstruction (GENRE)~\cite{price04} and its derived GEnome scale Model (GEM)~\cite{feist08}. This is important to differentiate between the established biochemical knowledge included in a GENRE and the modelling assumptions required for analysis or simulation with a GEM. A GENRE serves as a structured knowledge base of established biochemical facts, while a GEM is a model which supplements the established biochemical information with additional (potentially hypothetical) information to enable computational simulation and analysis~\cite{heavner12}. Reactions added to the GEM include the biomass objective function -- a sink representing cellular growth -- and hypothetical transporters.

Three versions of the network are made available:

\begin{itemize}
\item \texttt{<organism>\_<version>.xml}, a GEM for use in flux analyses, provided in Flux Balance Constraints (FBC) format~\cite{fbc}
\item \texttt{<organism>\_<version>\_cobra.xml}, the same GEM network, provided in Cobra format~\cite{schellenberger11}
\item \texttt{<organism>\_<version>\_recon.xml}, a GENRE containing only reactions for which there is experimental evidence
\end{itemize}

\section*{YeastNet}

YeastNet is an annotated metabolic network of Saccharomyces cerevisiae S288c that is periodically updated by a team of collaborators from various research groups. It started on the shoulders of previous reconstructions of the yeast metabolic network that were published separately (iLL672~\cite{kuepfer05} and iMM904~\cite{mo09}). However, due to the different approaches utilised, those earlier reconstructions had a significant number of differences. A community effort in 2007 resulted in a consensus network representation of yeast metabolism, reconciling the earlier results.

As of December 2012, six versions of the network have been released (see \tabref{3}).

\begin{table}[!ht]
\begin{center}
\begin{tabular}{| c | c | c |}
	\hline 
	version	& 	date				& 	publications		\\
	\hline 
	1		&	February 2008 		&	\cite{herrgard08}	\\
	2		&	June 2009 		&	--				\\
	3		&	October 2009 		&	--				\\
	4		&	March 2010 		&	\cite{dobson10}	\\
	5		&	September 2011 	&	\cite{heavner12}	\\
	6		&	December 2012 	&	--				\\
	\hline
\end{tabular}
\caption{Development of YeastNet}
\label{table:3}
\end{center}
\end{table}

The EcoliNet and YeastNet networks are structured identically to facilitate comparative studies. YeastNet is available from \href{http://yeast.sf.net/}{http://yeast.sf.net/}.

\paragraph{Acknowledgements}

This work is deliverable 4.1 of the EU FP7 (KBBE) grant 289434 ``\href{http://www.biopredyn.eu}{BioPreDyn}: New Bioinformatics Methods and Tools for Data-Driven Predictive Dynamic Modelling in Biotechnological Applications''.

\end{document}